**needLR: Long-read structural variant annotation with population-scale frequency estimation**

Jonas A. Gustafson[1,2,*], Jiadong Lin[3], Evan E. Eichler[3,4,5], Danny E. Miller[2,5,6,*]

1. Department of Molecular and Cellular Biology, University of Washington, Seattle, WA 98195, USA
2. Department of Pediatrics, University of Washington, Seattle, WA 98195, USA
3. Department of Genome Sciences, University of Washington School of Medicine, Seattle, WA 98195, USA
4. Howard Hughes Medical Institute, University of Washington, Seattle, WA, 98195, USA
5. Brotman Baty Institute for Precision Medicine, University of Washington, Seattle, WA 98195, USA
6. Department of Laboratory Medicine and Pathology, University of Washington, Seattle, WA 98195, USA

**\* Contact**: jgust1@uw.edu, dm1@uw.edu

**ABSTRACT**

**Summary:** We present needLR, a structural variant (SV) annotation tool that can be used for filtering and prioritization of candidate pathogenic SVs from long-read sequencing data using population allele frequencies, annotations for genomic context, and gene–phenotype associations. When using population data from 500 presumably healthy individuals to evaluate nine test cases with known pathogenic SVs, needLR assigned allele frequencies to over 97.5% of all detected SVs and reduced the average number of novel genic SVs to 121 per case while retaining all known pathogenic variants.

**Availability and Implementation:** needLR is implemented in bash with dependencies including Truvari v4.2.2, BEDTools v2.31.1, and BCFtools v1.19. Source code, documentation, and pre-computed population allele frequency data are freely available at https://github.com/jgust1/needLR under an MIT license.

**Contact**: jgust1@uw.edu, dm1@uw.edu

1. **INTRODUCTION**

More than half of individuals with a suspected genetic disorder remain undiagnosed after comprehensive clinical genetic testing (Dawood et al. 2025; Wojcik et al. 2023). A major contributor to this low diagnostic yield is the limited sensitivity of existing methods, such as short-read sequencing (SRS), for detecting or fully resolving structural variants (SVs; insertions, deletions, inversions, and duplications ≥50 base pairs) (Mandelker et al. 2016; Sudmant et al. 2015). While SVs represent a significant proportion of pathogenic variants underlying genetic disorders, SRS-based methods identify only about one-third to one-half of the total number of SVs per individual (Zhao et al. 2021; Chaisson et al. 2019). Long-read sequencing (LRS) is able to identify and fully resolve nearly all of the ~25,000 SVs per genome, providing a more comprehensive list of the SVs present in any one individual (Audano et al. 2019; Gustafson et al. 2024).

      While LRS can identify and resolve more SVs than SRS—increasing diagnostic yield among molecularly undiagnosed individuals—the filtering, prioritization, and interpretation of these variants has been challenging due to a limited amount of publicly available LRS-derived population data. To date, most publicly available population SV data have been generated using SRS and therefore contain only SRS-detectable SVs, leading to undercounting of population-level SVs and inaccurate variant annotations, particularly for large insertions (Collins et al. 2020; Dominguez Gonzalez et al. 2024). It is unknown whether SRS-derived SV calls can be used to accurately filter SVs detected by LRS.

Several efforts are underway to generate population-scale LRS data from diverse cohorts, including the Human Pangenome Reference Consortium (HPRC), Human Genome Structural Variation Consortium (HGSVC), the *All of Us* Research Program, work by Schloissnig and colleagues, and the 1000 Genomes Project Long-Read Sequencing Consortium (1KGP-LRSC) (Liao et al. 2023; Ebert et al. 2021; Garimella et al. 2025; Schloissnig et al. 2025; Gustafson et al. 2024). The 1KGP-LRSC has generated high coverage (>25×), high read length (N50 > 40 kbp) data on the Oxford Nanopore Technologies (ONT) platform for 500 individuals (1,000 haplotypes) from the 1KGP. Together, variant calls from these datasets are likely to provide more accurate SV population allele frequency estimates and more accurate cataloging of SVs, both of which will be essential for distinguishing pathogenic from benign SVs in clinical applications.

As LRS-based clinical applications are brought online and the amount of LRS-derived population-level SV data increases, there is a need for tools that can accurately and efficiently perform SV annotation, filtering, and prioritization (Garimella et al. 2025). Existing annotation tools provide either LRS-based allele frequency estimates without comprehensive genomic context or offer detailed functional annotation while relying on SRS-derived frequency databases (Geoffroy et al. 2018; Nicholas et al. 2022; Danis et al. 2022; Zheng et al. 2024). This fragmented landscape forces researchers to perform annotation using multiple tools followed by manual integration of results, creating bottlenecks in clinical variant interpretation workflows. To comprehensively prioritize candidate pathogenic SVs identified by LRS for clinical diagnosis, integrated tools that combine LRS-specific population frequencies with genomic and clinical context annotation are needed. Here, we present needLR, a comprehensive SV annotation pipeline specifically optimized for long-read SV calls, combining Truvari-based SV merging for allele frequency calculations with extensive genomic context annotation and candidate phenotype associations.

## 2. IMPLEMENTATION

### 2.1 needLR input

needLR requires three inputs: 1) a list of query VCFs, 2) a pre-merged population VCF, and 3) a reference genome used for alignment of both the query and population VCFs **(Figure 1)**. Users can choose between two population VCF options: a pre-merged VCF comprising 500 ONT samples from the 1KGP-LRSC, or their own custom VCF built for specific research needs. The pre-merged 1KGP-LRSC-derived population provides the advantage of a large, high-quality collection of diverse, presumably healthy individuals, while a custom VCF may be advantageous for users analyzing specific disease cohorts or understudied populations.

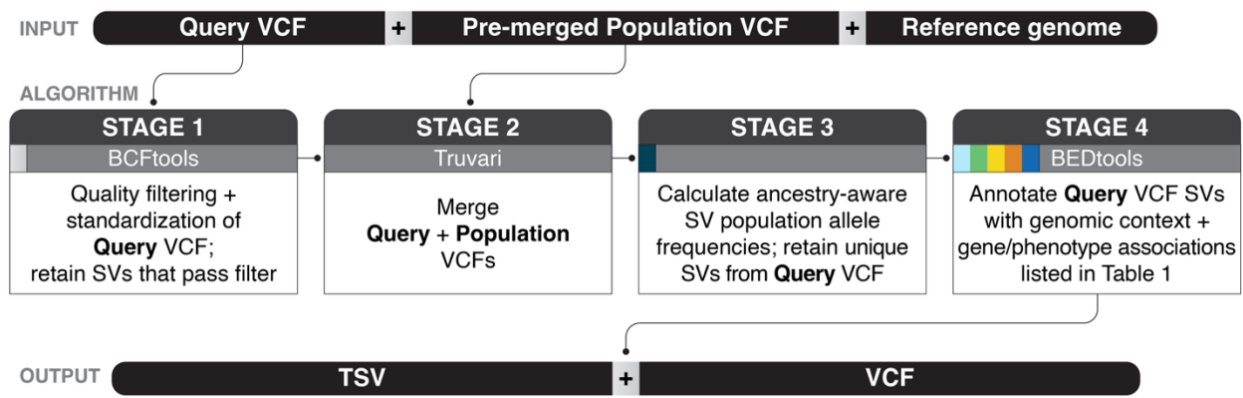

**Figure 1. needLR workflow.** needLR integrates a query VCF, a population control VCF (500 1kGP-LRSC samples by default), and a reference genome. The algorithm performs quality filtering, merging, ancestry-aware allele frequency estimation, and genomic and phenotypic annotation of structural variants. These data are output as both a tab-separated summary (TSV) and annotated VCF.

**2.2 Algorithm**

needLR uses a four-stage algorithmic approach that has been optimized for long-read SV annotation **(Figure 1)**. In Stage 1, quality filtering and standardization of query VCF files is performed using bcftools, retaining SVs ≥50 bp that pass the SV caller's internal filtering parameters (Filter=PASS) and are located on chromosomes 1–22, X, and Y, and the mitochondrial genome (Danecek et al. 2021). Stage 2 merges the query VCF with a pre-merged population VCF using Truvari v4.2.2 with user-defined matching parameters, including sequence similarity threshold, size similarity tolerance, and reference distance tolerance to account for long-read alignment variability (English et al. 2022). In Stage 3, ancestry-aware allele frequencies are calculated for each SV in the query VCF based on genotype counts from the population VCF. Finally, Stage 4 employs BEDTools and a backend directory of BED files to annotate each SV with genomic context and potential disease associations, enabling users to filter variants based on allele frequency, genomic context, and/or phenotype of interest (Quinlan and Hall 2010).

**Population allele frequency calculation.** Allele frequencies are calculated from genotype data extracted from a Truvari-merged population VCF. Both overall and ancestry-specific frequencies are computed across the five 1KGP superpopulations (African, American, East Asian, European, and South Asian) when using the needLR-provided population VCF. Quality control is implemented through Hardy-Weinberg equilibrium testing, which applies chi-square statistics to observed heterozygote and homozygote counts to flag variants with unexpected genotype distributions.

**Genomic context annotation.** needLR intersects SV coordinates with 12 comprehensive annotation tracks **(Table 1)** using BEDTools intersect (Mudge et al. 2025; Hamosh et al. 2000; Perez et al. 2025; Ren et al. 2023; Dwarshuis et al. 2024). These tracks include SV overlap with canonical genes and exons, repetitive elements (short tandem repeats, variable number tandem repeats, and other RepeatMasker-identified regions), segmental duplications, low-confidence and satellite-sequence regions (centromeres, pericentromeres, telomeres), and regions with defined sequence quality metrics (GIAB high-confidence regions and hg38 assembly gaps).

**Table 1. Sources of annotation data used by needLR**

| Annotation | Source |
| --- | --- |
| **Genes** | GENCODE v45[1] |
| **OMIM** | OMIM[2] |
| **Exonic** | Gencode v45[3] |
| **Centromeric** | UCSC table browser[4] |
| **Pericentromeric** | 5 Mbp added to either end of centromere coordinates[4] |
| **Telomeric** | 5 Mbp from chromosome ends (hg38)[5] |
| **STR** | Vamos[6] |
| **VNTR** | Vamos[6] |
| **Segdup** | GIAB Stratifications v3.3 (hg38) |
| **Repeat** | UCSC table browser[7] |
| **Gap** | UCSC table browser[8] |
| **HiConf** | GIAB DEFRABB[9] |

[1] http://ftp.ebi.ac.uk/pub/databases/gencode/Gencode_human/latest_release/gencode.v45.annotation.gtf.gz; (column 3 = gene | column 12 = protein_coding) + 1 kbp on each end.
[2] Accessed August 2023.
[3] http://ftp.ebi.ac.uk/pub/databases/gencode/Gencode_human/latest_release/gencode.v45.annotation.gtf.gz; Ensembl_canonical | column 3 = exon | gene name is in 3_GENE_PROTEIN_CODING_gencode.v45.annotation.bed (grep for protein_coding | column 3 = gene).
[4] Accessed 1/20/2024; Assembly: GRCh38, Group: Mapping and Sequencing, Track: Centromeres, Table: Centromeres.
[5] *https://hgdownload.cse.ucsc.edu/goldenpath/hg38/bigZips/hg38.chrom.sizes*
[6] https://zenodo.org/records/8357361/original_motifs.set148.bed.gz.
[7] Accessed 6/30/2024; Group: Repeats, Track: RepeatMAsker, table: rmsk.
[8] Accessed 1/20/2024; Assembly: GRCh38, Group: Mapping and Sequencing, Track: Gap, Table: Gap.
[9] https://ftp-trace.ncbi.nlm.nih.gov/ReferenceSamples/giab/data/AshkenazimTrio/analysis/NIST_HG002_DraftBenchmark_defrabbV0.012-20231107/GRCh38_HG002-T2TQ100-V1.0_stvar.benchmark.bed.

## 2.3 needLR output

The output from needLR includes basic information for each SV in the query VCF, including genomic coordinates, reference and alternate alleles, total supporting read counts, alternate allele supporting reads, reference allele supporting reads, SV length and type, and sample IDs of samples in the population VCF harboring the same SV. For SVs that match variants in the population VCF, the reported details correspond to the most frequently represented variant within that merged group. For SVs unique to the query VCF, the original variant details are retained. The final needLR output format is both a tab-separated TSV file and a VCF (v4.2) for downstream application compatibility.

## 2.4 Computational performance

All needLR processes run on a single thread. Using the precomputed database of 500 1KGP samples, needLR completes annotation of one input human genome sample in approximately 20 minutes.

## 3. VALIDATION

We validated needLR using nine positive control samples with known pathogenic SVs (5 deletions, 3 insertions, and 1 inversion) that were not identified by standard clinical SRS but were subsequently detected by LRS-based methods **(Table 2)**. Because Sniffles v2.5.2 identified all nine SVs in our positive control cohort, we selected it as the SV caller for needLR. To ensure consistency, individual VCFs that comprise needLR's provided backend control dataset population VCF (500 samples from the 1KGP-LRSC) were also generated using Sniffles v2.5.2.

**Table 2. Pathogenic SVs used for needLR validation.** needLR correctly annotated nine LRS-resolved pathogenic SVs as being unique to their respective individual when compared to SVs from 500 1KGP-LRSC samples.

| Sample | Pore | Sniffles2 calls | | | | | | Aligned BAM stats | | |
|---|---|---|---|---|---|---|---|---|---|---|
| | | SV type[1] | Gene(s) | Length (bp) | Chr | Start | End | Yield (Gbp) | Estimated coverage | N50 (kbp) |
| A | R9 | DEL | *FGA* | –4,147 | 4 | 154,589,944 | 154,594,091 | 91.4 | 28.6 | 31.7 |
| B | R9 | DEL | *FANCD2* | –459 | 3 | 10,049,302 | 10,049,761 | 44.2 | 13.8 | 32.6 |
| C | R10 | INS | *ABCD1* | 2,731 | X | 153,731,828 | 153,731,828 | 56.5 | 17.6 | 10.2 |
| D | R9 | DEL | *IKBKG* | –2,356 | X | 154,563,099 | 154,565,455 | 104.2 | 34.0 | 5.8 |
| E | R10 | DEL | *DBT* | –3,111 | 1 | 100,218,762 | 100,221,873 | 62.3 | 19.5 | 5.8 |
| F | R9 | DEL | *AGRN* | –1,210 | 1 | 1,041,232 | 1,042,442 | 118.8 | 37.1 | 45.1 |
| G | R10 | INV | *COL5A1* | 1,470,536 | 9 | 134,789,327 | 136,259,863 | 62.3 | 19.5 | 5.8 |
| H | R10 | INS | *AGL* | 319 | 1 | 99,916,500 | 99,916,500 | 112.4 | 35.1 | 5.1 |
| I | R10 | INS | *ASPA* | 2,622 | 17 | 3,490,218 | 3,490,218 | 117.6 | 38.0 | 23.2 |

[1] DEL, deletion; INS, insertion; INV, inversion.

Using this Truvari-merged population VCF, needLR correctly annotated all pathogenic variants as unique to their respective individuals (i.e., absent from the population VCF), confirming its ability to accurately annotate rare SVs. With Truvari parameters optimized for this application, needLR removed >97% of common SVs (AF > 0) across the 9 samples, filtering average SVs per individual from 21,176 total SVs to 483 unique SVs **(Figure 2A)**, and further prioritized unique SVs based on their genomic context **(Figure 2B)**. The number of first-pass putative candidate pathogenic SVs (unique SVs that intersect an exon of an OMIM-associated gene) was reduced to an average of only 3 per individual (**Figure 2C**). In comparison, applying the same Truvari parameters with gnomAD v4.1 (63,046 unrelated short-read genomes) as the reference database filtered only 44% of SVs, leaving an average of 11,756 total unique SVs and 26 OMIM-associated exonic SVs per individual.

## 3.1 Comparative performance analysis

Several tools exist to prioritize candidate pathogenic SVs based on allele frequency, genomic context, disease association, or combinations thereof. We compared the performance of four tools with utility closest to needLR—SVAFotate, STIX, SvAnna, and AnnotSV—on our nine positive control samples.

SVAFotate compares the coordinates and SV type from a query VCF to a backend BED file generated from publicly available SRS-based SV calls. Using the recommended parameter of 0.8 reciprocal overlap, SVAFotate filtered an average of 34.9% of SVs per positive control sample as present in public datasets and annotated 8/9 known pathogenic SVs as unique to the affected individual. SVAFotate does not provide genomic context or disease association information.

STIX indexes and creates a database of all SV evidence from a cohort of BAM files to assign population allele frequencies to query SVs. We created a STIX database from the same 500 1KGP-LRSC samples used in needLR. STIX identified 5/9 known pathogenic SVs as unique to the affected individual. Like SVAFotate, STIX does not provide genomic context or disease association information.

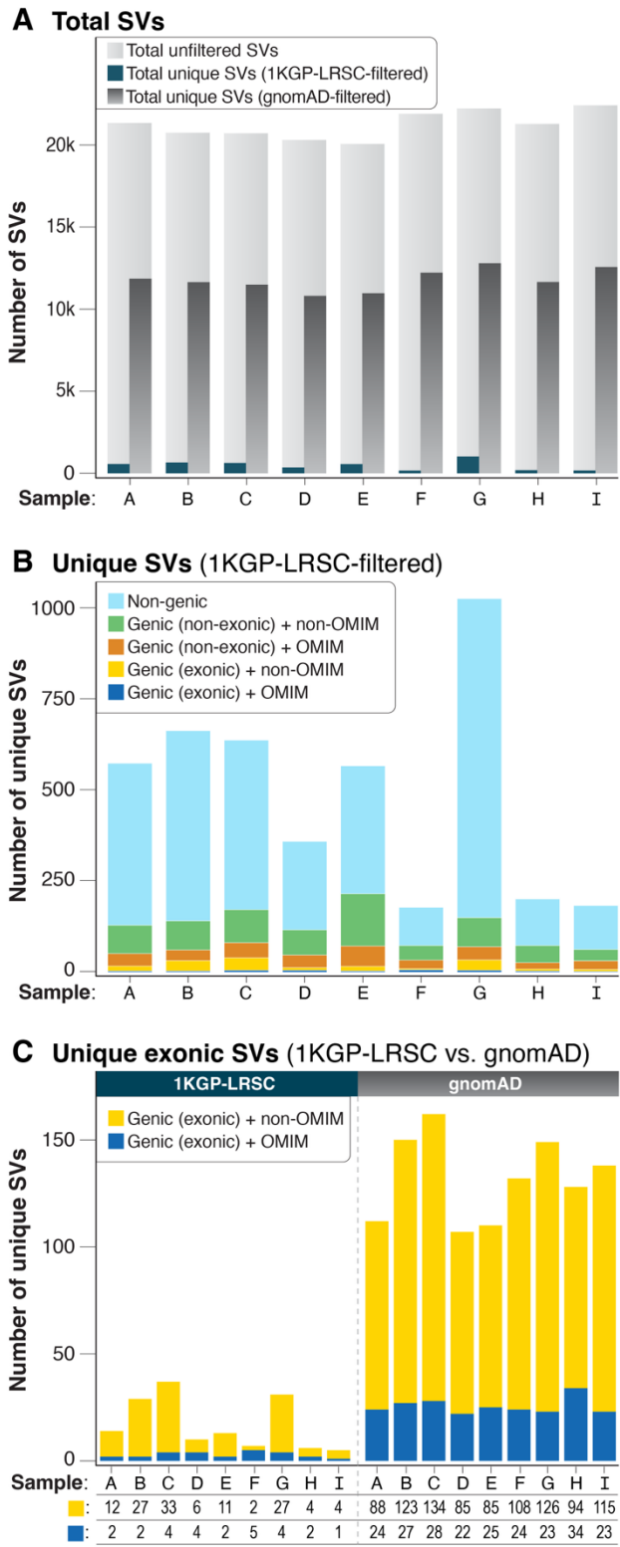

**Figure 2. needLR validation using 9 test cases**. (A) Running needLR using its default 1kGP-LRSC population VCF filtered out >97% of common SVs from test samples A–I. (B) needLR further categorizes and prioritizes remaining unique SVs based on genomic context (e.g., genic, exonic, OMIM-associated). (C) More exonic SVs are filtered out using the 1KGP-LRSC samples as the control population than when using gnomAD v4.1.

SvAnna requires both a query VCF and associated Human Phenotype Ontology (HPO) terms as input, necessitating *a priori* phenotype information. It filters SVs with minor allele frequency (MAF) >1% and >80% reciprocal overlap with common variants in source databases (including publicly available SRS and LRS data), then outputs SVs with calculated Pathogenicity of Structural Variation (PSV) scores based on the intersection of SVs with annotated genic regions and the association of affected genes with input HPO terms. SvAnna ranked 5/9 positive control pathogenic SVs with the highest PSV score, 2/9 with the second-highest score, and failed to rank 2/9 in the top 100 variants.

AnnotSV assigns ACMG-compatible pathogenicity scores based on a comprehensive database of SRS-based population allele frequencies, genomic contexts, and disease associations. Using default parameters without HPO terms, AnnotSV classified 3/9 positive control SVs as "likely pathogenic," 2/9 as "variant of unknown significance" (VUS) and did not assign pathogenicity scores to 4/9 samples.

## 4. DISCUSSION

needLR addresses critical gaps in long-read SV annotation through three key innovations: (1) integration of population-scale LRS data to provide LRS-derived allele frequency estimates, (2) customizable Truvari-based merging parameters for matching and merging query SVs with population data, and (3) comprehensive genomic context annotation enabling clinical-grade variant interpretation. This merging strategy offers computational efficiency advantages while maintaining annotation accuracy. Pre-computed population backends enable rapid query processing without sacrificing the benefits of population-scale frequency estimation, making needLR suitable for both research and clinical applications (Sui et al. 2025).

We show that using needLR with an LRS-derived control SV database (500 1KGP-LRSC genomes) reduces the number of candidate pathogenic SVs by almost an order of magnitude as compared to using the SRS-derived control SV data from gnomAD v4.1. This trend extends to clinically relevant regions such as OMIM-associated genes and exons. The majority of SVs remaining after gnomAD filtering are insertions, which aligns with the established limitation of SRS in detecting insertions.

Current limitations of needLR include the exclusion of breakend variants and SVs >1 Mbp, which are most often false positive calls in our data. Sex chromosome analysis is limited by the ability of Sniffles2 to accurately distinguish sex chromosome alleles (i.e., a substantial number of Y-chromosome calls in XX individuals and heterozygous X-chromosome calls in XY individuals). Future developments will incorporate additional SV caller integration (cuteSV, SVIM, pbsv, etc.), trio analysis functionality for *de novo* variant detection, compatibility with new reference genomes (e.g., T2T-CHM13), technology-specific population backends from 1KGP samples, and machine-learning-based pathogenicity prediction leveraging the comprehensive annotation features.

A major goal of needLR is to enable clinical laboratories to leverage the advantages of LRS while maintaining standardized annotation workflows. The tool's ancestry-specific frequency calculations support diverse patient populations, while comprehensive genomic context annotation facilitates ACMG/AMP guideline implementation for SV interpretation.

needLR provides the first comprehensive annotation solution specifically designed for long-read SV annotation and analysis, addressing fundamental limitations of existing tools developed for short-read data. By leveraging population-scale LRS data and optimized algorithmic approaches, needLR enables accurate frequency estimation and clinical interpretation of SVs detected by LRS technologies. The tool's open-source availability, computational efficiency, and comprehensive annotation capabilities make needLR an essential component of both research and clinical bioinformatics pipelines.

**ACKNOWLEDGEMENTS**
We thank Angela Miller for assistance with manuscript preparation.

**Funding**
This work was supported by the National Institutes of Health through the NIH Director's Early Independence Award DP5OD033357 to D.E.M. Research reported in this publication was supported, in part, by the National Human Genome Research Institute of the National Institutes of Health (NIH) under Award Number R01HG010169 to E.E.E. The content is solely the responsibility of the authors and does not necessarily represent the official views of the NIH. E.E.E. is an investigator of the Howard Hughes Medical Institute.

**Author Contributions**
J.A.G.: Conceptualization, Data Curation, Formal analysis, Software, Writing- original draft, Writing – review & editing
J.L.: Data Curation, Formal analysis, Writing – review & editing
E.E.E.: Resources, Methodology, Writing – review & editing
D.E.M.: Resources, Methodology, Writing – original draft, Writing – review & editing

**Disclosures**
J.A.G. has received travel support from ONT. D.E.M. is on scientific advisory boards at ONT, Inso Biosciences, and Basis Genetics; is engaged in research agreements with ONT and PacBio; has received research and travel support from ONT and PacBio; holds stock options in

**Data Availability**

Source code, documentation, and pre-computed population allele frequency data are freely available at https://github.com/jgust1/needLR under an MIT license.